\documentclass[twocolumn]{aastex63}
\usepackage{amsmath}
\usepackage{amssymb}
\usepackage{multirow}

\newcommand{\fesc}{\ifmmode{f_{\rm esc}}\else{$f_{\rm esc}$}\fi}
\newcommand{\fescs}{\ifmmode{f_{\rm esc}^\star}\else{$f_{\rm esc}^\star$}\fi}
\newcommand{\kms}{\ifmmode{{\;\rm km~s^{-1}}}\else{km~s$^{-1}$}\fi}
\newcommand{\fgas}{\ifmmode{{f_{\rm gas}}}\else{$f_{\rm gas}$}\fi}
\newcommand{\cubecm}{\ifmmode{{\rm cm^{-3}}}\else{cm$^{-3}$}\fi}
\newcommand{\ztwo}{\ifmmode{{\rm [Z_2/H]}}\else{[Z$_2$/H]}\fi}
\newcommand{\zthree}{\ifmmode{{\rm [Z_3/H]}}\else{[Z$_3$/H]}\fi}
\newcommand{\lsim}{\lower0.3em\hbox{$\,\buildrel <\over\sim\,$}}
\newcommand{\gsim}{\lower0.3em\hbox{$\,\buildrel >\over\sim\,$}}

\newcommand{\sfr}{\ifmmode{\textrm{M}_\odot \,\textrm{yr}^{-1} \,\textrm{Mpc}^{-3}}\else{M$_\odot$ yr$^{-1}$ Mpc$^{-3}$}\fi}
\newcommand{\hsfr}{\ifmmode{\textrm{M}_\odot\, \textrm{yr}^{-1}}\else{M$_\odot$ yr$^{-1}$}\fi}

\newcommand{\eavg}{\ifmmode{\langle E_\gamma \rangle}\else{$\langle E_\gamma \rangle$}\fi}

\newcommand{\Ms}{\ifmmode{M_\odot}\else{$M_\odot$}\fi}
\newcommand{\vrms}{\ifmmode{v_{\rm rms}}\else{$v_{\rm rms}$}\fi}

\newcommand{\tvir}{\ifmmode{T_{\rm{vir}}}\else{$T_{\rm{vir}}$}\fi}
\newcommand{\mvir}{\ifmmode{M_{\rm{vir}}}\else{$M_{\rm{vir}}$}\fi}
\newcommand{\rvir}{\ifmmode{r_{\rm{vir}}}\else{$r_{\rm{vir}}$}\fi}

\newcommand{\jj}{\ifmmode{J_{21}}\else{$J_{21}$}\fi}
\newcommand{\flw}{\ifmmode{F_{LW}}\else{$F_{LW}$}\fi}
\newcommand{\kph}{\ifmmode{k_{\rm ph}}\else{$k_{\rm ph}$}\fi}

\newcommand{\zsun}{\ifmmode{\rm\,Z_\odot}\else{$\rm\,Z_\odot$}\fi}

\newcommand{\hi}{H {\sc i}}
\newcommand{\hii}{H {\sc ii}}
\newcommand{\hei}{He {\sc i}}
\newcommand{\heii}{He {\sc ii}}
\newcommand{\heiii}{He {\sc iii}}
\newcommand{\nhi}{\ifmmode{N_{\rm HI}}\else{$N_{\rm HI}$}\fi}

\begin{document}

\title{The Galaxy Assembly and Interaction Neural Networks ({\sc GAINN}) for high-redshift JWST observations}

\correspondingauthor{Lillian Santos-Olmsted}
\email{solmsted@stanford.edu}

\author[0000-0002-8763-3702]{Lillian Santos-Olmsted}
\affiliation{Kavli Institute for Particle Astrophysics and Cosmology, Stanford University, Stanford, CA 94305, USA}

\author[0000-0002-8638-1697]{Kirk S.~S.~Barrow}
\affiliation{Department of Astronomy, University of Illinois at Urbana-Champaign, 
1002 W Green St, Urbana, IL 61801, USA}

\author[0000-0001-6742-8843]{Tilman Hartwig}
\affiliation{Department of Physics, School of Science, The University of Tokyo, Bunkyo, Tokyo 113-0033, Japan}
\affiliation{Institute for Physics of Intelligence, School of Science, The University of Tokyo, Bunkyo, Tokyo 113-0033, Japan}
\affiliation{Kavli Institute for the Physics and Mathematics of the Universe (WPI), The University of Tokyo Institutes for Advanced Study, The University of Tokyo, Kashiwa, Chiba 277-8583, Japan}

\begin{abstract}
We present the Galaxy Assembly and Interaction Neural Networks ({\sc Gainn}), a series of artificial neural networks for predicting the redshift, stellar mass, halo mass, and mass-weighted age of simulated galaxies based on JWST photometry. Our goal is to determine the best neural network for predicting these variables at $11.5 < z < 15$. The parameters of the optimal neural network can then be used to estimate these variables for real, observed galaxies. The inputs of the neural networks are JWST filter magnitudes of a subset of five broadband filters (F150W, F200W, F277W, F356W, and F444W) and two medium-band filters (F162M and F182M). We compare the performance of the neural networks using different combinations of these filters, as well as different activation functions and numbers of layers. The best neural network predicted redshift with normalized root mean squared error NRMS = $0.009_{-0.002}^{+0.003}$, stellar mass with RMS = $0.073_{-0.008}^{+0.017}$, halo mass with MSE = $ 0.022_{-0.004}^{+0.006}$, and mass-weighted age with RMS = $10.866_{-1.410}^{+3.189}$. We also test the performance of {\sc Gainn} on real data from MACS0647—JD, an object observed by JWST. Predictions from {\sc Gainn} for the first projection of the object (JD1) have mean absolute errors $\langle \Delta z \rangle <0.00228$, which is significantly smaller than with template-fitting methods. We find that the optimal filter combination is F277W, F356W, F162M, and F182M when considering both theoretical accuracy and observational resources from JWST. 
\end{abstract}

\section{Introduction}
Accurate photometric redshift (photo-z) estimates are essential for knowing distances to very far objects in the universe and can therefore indirectly help astronomers learn more about other galactic variables, including stellar mass, halo mass, and mass-weighted age. Consequently, photo-z estimates are crucial for investigating galactic properties and evolution, as well as cosmology \citep{Fontana1999,Hu_1999,Salvato2019}. The determination of redshift from photometric measurements was initially explored several decades ago \citep{Baum1957,Baum_1962}, as an alternative to spectroscopic measurements, which are difficult to obtain for faint sources that require long exposure times. This approach of using filter magnitudes to estimate redshift was solidified in later works \citep{Butchins_1981,Koo_1985,Connolly_1995,Hogg_1998}. 

Photo-z predictions have been made using a variety of methods, which can generally be classified into one of two overarching categories: Spectral Energy Distribution (SED) template-fitting methods and Machine Learning (ML) methods. In SED template-fitting methods, the spectra of galaxies are obtained from a library of template spectra derived from empirical data \citep{Coleman_1980,Kinney_1996} or synthetic data \citep{Fioc_1997,Bruzual2003,Maraston_2005,Conroy2010}. These template spectra are converted into fluxes, taking into account the transmission curves of the chosen filters. The template fluxes are then compared to the fluxes of observed galaxies, in order to determine which template minimizes the error between the observed and template fluxes \citep{Fontana_2000}. This comparison is often performed through a $\chi^2$ fit using existing software (e.g. {\sc Le Phare} \citep{Arnouts_2011}, {\sc Hyperz} \citep{Bolzonella_2011}). Once the optimal template is found, the corresponding redshift can be predicted. There have been many codes developed recently which undergo the entire process of predicting photo-z, from template spectra to redshift (e.g. \textsc{Bagpipes} \citep{2018MNRAS.480.4379C}, \textsc{piXedfit} \citep{Abdurrouf2021}, \textsc{BEAGLE} \citep{Chevallard2016_BEAGLE}, \textsc{Prospector} \citep{Leja2017,Johnson2021}, and \textsc{EASY} \citep{Brammer2008}). This method has been widely applied in the recovery of redshift and other properties using filter fluxes from the James Webb Space Telescope (JWST) \citep{Bisigello12016,Bisigello_2017,Kemp_2019,Roberts_Borsani_2021}.

Machine learning methods estimate redshift by considering a data set of known magnitudes and redshifts and learning the features that relate them. These learned relationships can then be used to predict redshift in a blind data set where nothing is known about the true redshift values. As in SED template-fitting, the goal of machine learning is to minimize the error between a predicted and true value, but in this case the value is the redshift itself and not a set of fluxes. Unlike SED template-fitting, ML algorithms calculate the errors of a single data set over multiple iterations and attempt to reduce the errors in subsequent iterations. Machine learning methods can be used in the same manner to recover other galaxy properties \citep{Calderon_2019,Simet_2019,Surana_2020}.

Provided there is a high-quality data set to train on, these ML techniques are applicable. The training data for observed redshifts generally comes from spectroscopic data. It has been shown that ML performs better than SED template-fitting at lower redshifts but worse than it at high redshifts, because there are not enough high-quality, spectroscopic observations of high redshift galaxies to provide a sufficiently large dataset for ML algorithms to train on \citep{Hildebrandt_2010, Salvato2019}. Previous studies using ML to estimate photometric redshifts have generally been restricted to low-redshift ranges up to $z \approx 4$ \citep{Firth_2003,Vanzella_2004,Almosallam_2015,Pasquet_2018,Razim_2021,Lee_2021,Lee_2022,Zhou_2022a,Zhou_2022b}. 

However, this barrier can be removed by using synthetic data to provide high redshift values for the ML algorithm to train on. One source of synthetic data are the spectral templates described earlier, but naturally they are based off existing data. In other words, a ML model training on template spectra would not be exposed to novel or unexpected features that would prepare it to handle real data. An alternative source of synthetic data are cosmological hydrodynamic simulations \citep{Hopkins_2014,Vogelsberger_2014,Genel_2014,Schaye_2014,Crain_2015,Xu_2016,Barrow_2020}, numerical models of the universe that simulate star formation, radiative transfer, and other dynamic physical phenomena based on a set of initial conditions \citep{Barrow_2017}. These simulations aim to model the evolution of galaxies from fundamental physics principles. To the extent of our knowledge of cosmology and access to computational resources, the properties of galaxies within the simulation are determined from the same physical laws that govern the behavior of real galaxies. Therefore, cosmological simulations can provide realistic data sets at high redshift for ML algorithms to learn from. In addition, the time-dependent aspect of cosmological simulations supplies more data points by allowing users to record a galaxy's properties at different snapshots in time.
Data from cosmological simulations has already been used to train machine learning models that predict the stellar mass and halo mass of simulated galaxies \citep{Kamdar_2016,Agarwal_2018,Gilda_2021,Villanueva-Domingo_2022,Zhou2022}, in addition to other observables \citep{Villaescusa-Navarro_2021a, Villaescusa-Navarro_2021b,Lovell_2019}. 

For this approach to be useful for predicting the redshifts and other properties of real galaxies, the ML algorithm needs to optimize a mathematical relationship between the magnitudes and redshifts of simulated galaxies that is directly transferable to observed galaxies. This relationship would be easier to investigate in surveys of high-redshift galaxies, because cosmological simulations start near the beginning of the universe and require less computing time to reach higher redshifts. Conveniently, the recent launch of JWST will transform our ability to study the properties of high redshift galaxies and allow us to probe the very early universe. Of the four main instruments on JWST, the one that is of most interest in this work is the Near-Infrared Camera (NIRCam). NIRCam includes filter imaging through a short wavelength channel (0.6 to 2.3 microns) and a long wavelength channel (2.4 to 5.0 microns) \citep{Gardner2006}. The various broad and medium-band filters on NIRCam provide an optimal wavelength range to perform photometric measurements of redshift and other properties at $11.5 \lesssim z \lesssim 15$. 

Several candidate galaxies in this redshift range have already been observed by NIRCam (e.g. \citep{Adams_2022,Bradley2022,Castellano_2022,Donnan_2023,Harikane_2023}). Some high-redshift galaxy candidates recently observed by JWST have higher stellar masses than predicted by the standard cosmological model ($\Lambda$CDM) \citep{Boylan-Kolchin_2022,labbe_2022,Menci_2022}. According to $\Lambda$CDM, there is an upper bound on the number density and stellar mass density of galaxies as a function of stellar mass and redshift. Since there are candidate galaxies at the edge of the bound, these observations could potentially challenge the standard cosmological model, depending on the accuracy of the measured redshifts and stellar masses. Therefore, the accurate determination of these variables is instrumental in evaluating the validity of this challenge to $\Lambda$CDM.

In this work, we present Galaxy Assembly and Interaction Neural Networks ({\sc Gainn}), a set of artificial neural networks used to recover the photometric redshifts of galaxies from cosmological simulations. It is assumed that objects studied using predictions from {\sc Gainn} have already been identified as high redshift galaxies in our redshift range. We use different combinations of five broadband (F150W, F200W, F277W, F356W, F444W) and two medium-band (F162M, F182M) JWST filters. We also use {\sc Gainn} to recover other properties of the galaxies: stellar mass, halo mass, and mass-weighted age. The inputs of the neural networks (NNs) are explored as a parameter space to optimize performance and minimize error. We compare the performance of various multi-layered NNs using the machine learning software {\sc Tensorflow} v2.4.1 \citep{tensorflow}. To account for the various exposure times for different JWST filters, we address three potential strategies: 1) The best outcomes that theory can offer with minimal observational constraints; 2) A strategy which incorporates both theoretical optimism and observational feasibility; and 3) The most ideal strategy given JWST resources. 

This paper is structured as follows: In sections \ref{cosmoSim} and \ref{radTransfer}, we explain the mechanics of our hydrodynamic cosmological simulation. We show the conversion between simulated galaxy spectra and filter magnitudes in section \ref{mag}. Sections \ref{NN} and \ref{params} include a discussion of how NNs function and our parameter selection for the NNs. Observational constraints are considered in section \ref{constraints}. We present our results for the unconstrained, observationally feasible, and observationally ideal best-case scenarios in sections \ref{theo_results}, \ref{interm_results}, and \ref{obs_results}, respectively. In section \ref{testcase}, we present results from the first test of our NNs on real data. Finally, we summarize and discuss our results in section \ref{disc}. 

\section{Methods}
\subsection{Cosmological Simulation}
\label{cosmoSim}

The cosmological simulation is evolved using the radiation-hydrodynamic adaptive mesh refinement code {\sc Enzo}  \citep{Bryan2014}.  The initial conditions define a zoom-in region for the Lagrangian volume containing a $\sim 10^{11} \rm{M_{\odot}}$ halo at $z=0$ with a quiescent merger history as taken from a public repository provided by the  {\sc Agora} collaboration \citep{2014ApJS..210...14K}.  Similarly to its antecedent simulation in \citet{Barrow_2020},  the cosmological parameters of $\Omega_{M,0} = 0.3065$, $\Omega_{\Lambda,0} = 0.6935 $, $\Omega_{b,0} = 0.0483$, $h = 0.679$, $\sigma_8 = 0.8344$, and $n = 0.9681$ are used,  which are taken from the most recent release of the \citet{2020A&A...641A...6P}.  The root grid for both dark matter particles and the baryon mesh is a 60 Mpc$^3$ box containing 512$^3$ particles and cells respectively.  The zoom-in region is bounded by the dimensions $2578.12 \times 3750 \times 2109.38$ co-moving kpc$^3/ h^{-3}$ using four additional levels of nested refinement.  This results in a maximum dark matter resolution of $2.81 \times 10^4\ \rm{M}_\odot$ and an effective grid size of 8912$^3$.

Within the zoom-in region, ten to eleven more levels of gas refinement are allowed, resulting in cell sizes as small as a quarter of a proper parsec.  The star formation,  9-species (\hi, \hii, \hei, \heii, \heiii, e$^-$, H$_2$, H$_2^+$, H$^-$) primordial gas evolution,  and metal enrichment routines used are identical to \citep{Barrow_2020},  except that the minimum mass for metal enriched star clusters has been reduced to 500 $\rm{M}_\odot$.  By carefully monitoring and tracking the evolution of the stellar history in multiple test simulations, it was determined that star formation converged at these resolutions and minimum cluster masses.  Further levels of refinement were found to be counter-productive, resulting in diverging quantities of radiating star particles. Additionally, few metal-enriched star particles were produced at the minimum mass, effectively avoiding issues with the stochasticity of bright massive stars. We explicitly follow the formation of metal-free stars, although their contribution to the bolometric luminosity might not be significant for the first galaxies \citep{riaz_22}.

The cosmological simulation terminates at $z=11.36$ and includes the merger history of approximately 180 galaxies, saved as output data in 1 Myr intervals beginning right before the onset of star formation at $z >20$.  

\subsection{Radiative Transfer Post-Processing Routine Improvements}
\label{radTransfer}

\begin{figure*}[ht]
    \centering
    \includegraphics[width=15cm]{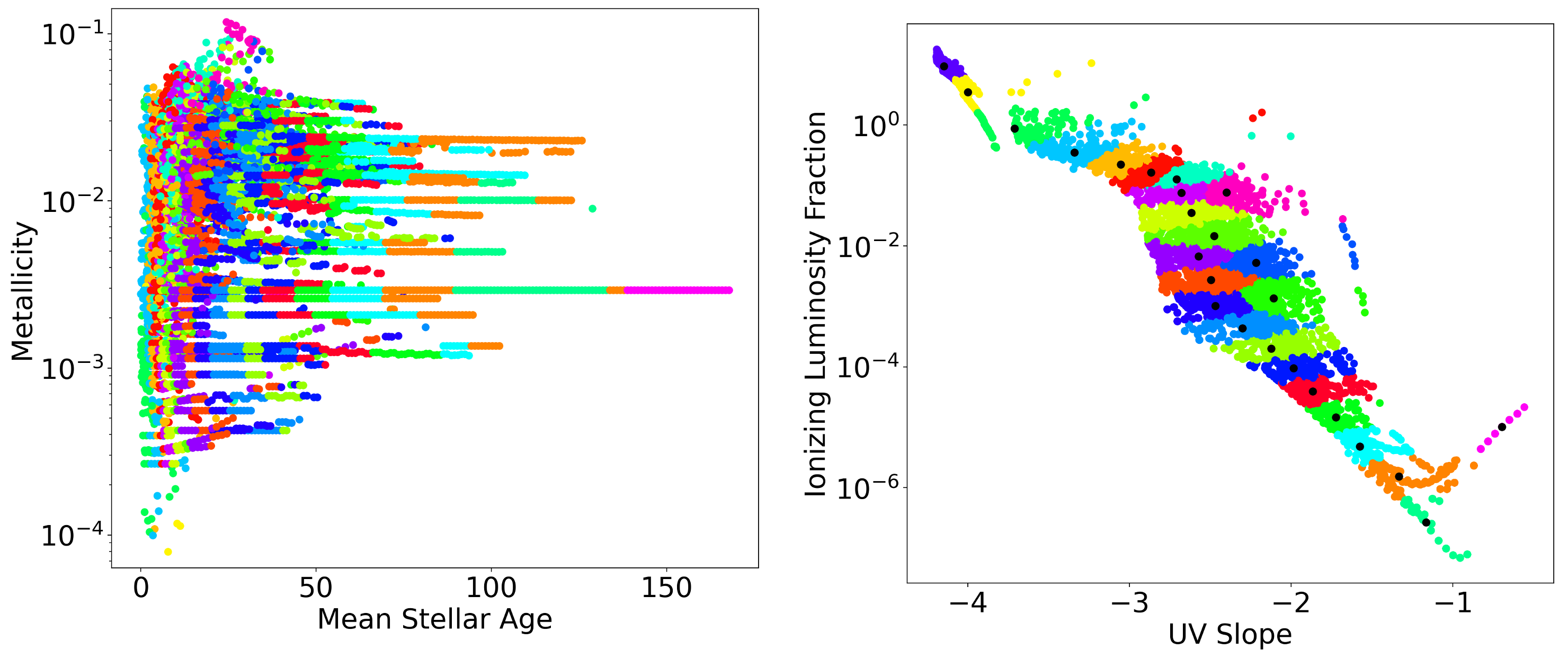}
    \caption{Left: Our sample of galaxies plotted by mean stellar age and metallicity showing tracks representing the time-evolution of star-forming halos. Right: The same plotted by the relationship between the fraction of ionizing to non-ionizing radiation and the UV slope, which serves as a proxy for spectral shape. Both figures are colored by k-means clustering performed on the distribution in the right figure. The archetypal galaxies in each cluster are colored with black markers.}
    \label{spectra_sample}
\end{figure*}

A radiative transfer post-processing routine is used to produce detailed spectra from simulation data. The methods employed are identical to those discussed in \citep{Barrow_2020} except that absorption look-up tables were produced using archetypal background galactic spectra instead of being recalculated at every timestep at every halo. This change was made to improve the scalability to the technique so that it can accommodate a larger number of merger histories.  

Mean intrinsic galactic continuum spectra are calculated using Flexible Stellar Population Synthesis \citep[FSPS;][]{Conroy2010} for metal-enriched populations and \citep[Yggdrasil;][]{Zackrisson_2011} for metal-poor stars using 55 $M_\odot$ as a cut off between using  PopIII.1 or PopIII.2 (top-heavy) models.  These spectra are then organized by the fraction of the luminosity in ionizing radiation and the UV slope. From these two variables, 25 archetypical background spectra are selected using k-means clustering to create photo-ionization sensitive temperature,  wavelength,  and metallicity look-up tables for gas absorption per nucleon density (see Fig. \ref{spectra_sample}).  Though stellar population synthesis models like {\sc FSPS} take metallicity and age as inputs for spectra shape, ionizing radiation and UV slope were found to much more naturally segregate the spectra while minimizing degneracies in resulting spectra.  

All the other elements of the radiative transfer pipeline are functionally unchanged.  To summarize, these absorption profiles are used for the ray tracing calculations from each source to each voxel of gas within the halo,  which are used to calculate the strength of emission lines throughout the halo and then used again to ray trace the final spectra of the galaxy at the virial radius, which can be a strong function of the orientation of the galaxy with respect to the observer's location.  For this study, the mean resulting spectra is used to create over 10,000 galaxy-timestep spectra. 

Early Universe galaxies typically have short, bursty star formation histories and so their spectra are highly sensitive to the timing and height of peaks in the star formation rate. The evolution of even a single galaxy can be used to model a large parameter space of output spectra by separately examining periods of a galaxy's star formation history, which represent a several possible combinations of star formation timings.

When taken together, the modeled spectra of the more than 10,000 galaxy-timesteps samples a diverse parameter space of spectral shape (Fig. \ref{spectra_sample}), which, along with redshift are key to disentangling galactic properties from observed filter fluxes. Thus, while we have a modest sample of galaxies in our final redshift, our time resolution and methodology provide an unprecedentedly large sample of synthetic spectra from robost, high-resolution radiative-hydrodynamic simulations and radiative transfer modeling. However, feedback prescriptions are still formulated based on best-effort theoretical arguments rather than empirically due to still-forthcoming observations of the early Reionization Universe so we caution that our results may incorporate unquantifiable theoretical biases, especially in our recovery of mean stellar age and halo mass. Additionally, more simulations will be incorporated into the model to extend the scope of our analysis in the near future so {\sc Gainn} should be seen as a continuously improving framework, not a final result.

\subsection{Filter Magnitudes}
\label{mag}
The spectra of the galaxies are obtained from the simulation in the form of energy values $L_{\nu}$ which each correspond to an emitted frequency $\nu_e$. The data for a given telescope filter $i$ consists of throughputs $R_i$, each corresponding to a specific observed wavelength. Each emitted wavelength is converted to an observed wavelength in order to map the throughputs to their corresponding emitted frequencies. Emitted spectral energy density blueward of the Lyman-$\alpha$ line (1215.67 \AA) is assumed to be zero due to pre-Reionization absorption by neutral hydrogen in the intergalactic medium. The observed flux $f_i$ through the filter is:
\begin{equation}
    f_i = \frac{1}{4\pi d_L^2} \int_{0}^{\infty} \frac{L_{\nu}(\nu_e)R_i(\nu_e)}{\nu_e} \ d\nu_e,
\end{equation}
where $d_L$ is the luminosity distance defined by 
\begin{equation}
    d_L = \frac{c(1+z)}{H_0} \int_{0}^{z} \frac{dz'}{\sqrt{\Omega_{M,0}(1+z')^3+\Omega_{\Lambda,0}}},
\end{equation}
where $H_0$, $c$, $\Omega_{\Lambda,0}$, and $\Omega_{M,0}$ have their usual cosmological definitions and are equivalent to the values used in the simulation. The AB magnitude $m_{AB,i}$ through filter $i$ is calculated from the flux as 
\begin{equation}
    m_{AB,i} = -2.5\log_{10}\frac{f_i}{\int_0^\infty \frac{R_{i}(\nu)}{\nu} d\nu} - 48.6,
\end{equation}
where $\nu$ is the observed frequency. 

\subsection{Neural Network}
\label{NN}
We explore an approach for determining the weights of the filter magnitudes by employing NNs. A neural network is widely known as an algorithm that can learn about the features of a system by training with examples and then apply that learning to new situations and real data. The primary advantage of NNs over the standard linear regression model is that some NN architectures are more adept at identifying nonlinear relationships. In principle, a basic NN consists of an input layer, an output layer, and usually one or more hidden layers. Each layer contains a set of neurons that are connected to each neuron from the previous layer and subsequent layer by a weight that scales the data and a bias that translates the data \citep{Schmidhuber_2015}. From one layer to the next, the values in each neuron are multiplied by the corresponding weight and added to the corresponding bias. The resulting values undergo a transformation through an activation function, and these transformed values then become the input for the following layer. This process is repeated until the NN generates an output from the final layer (see Eq. \ref{eq:multilayNN}). 

\begin{figure}[ht]
    \centering
    \includegraphics[width=8.5cm]{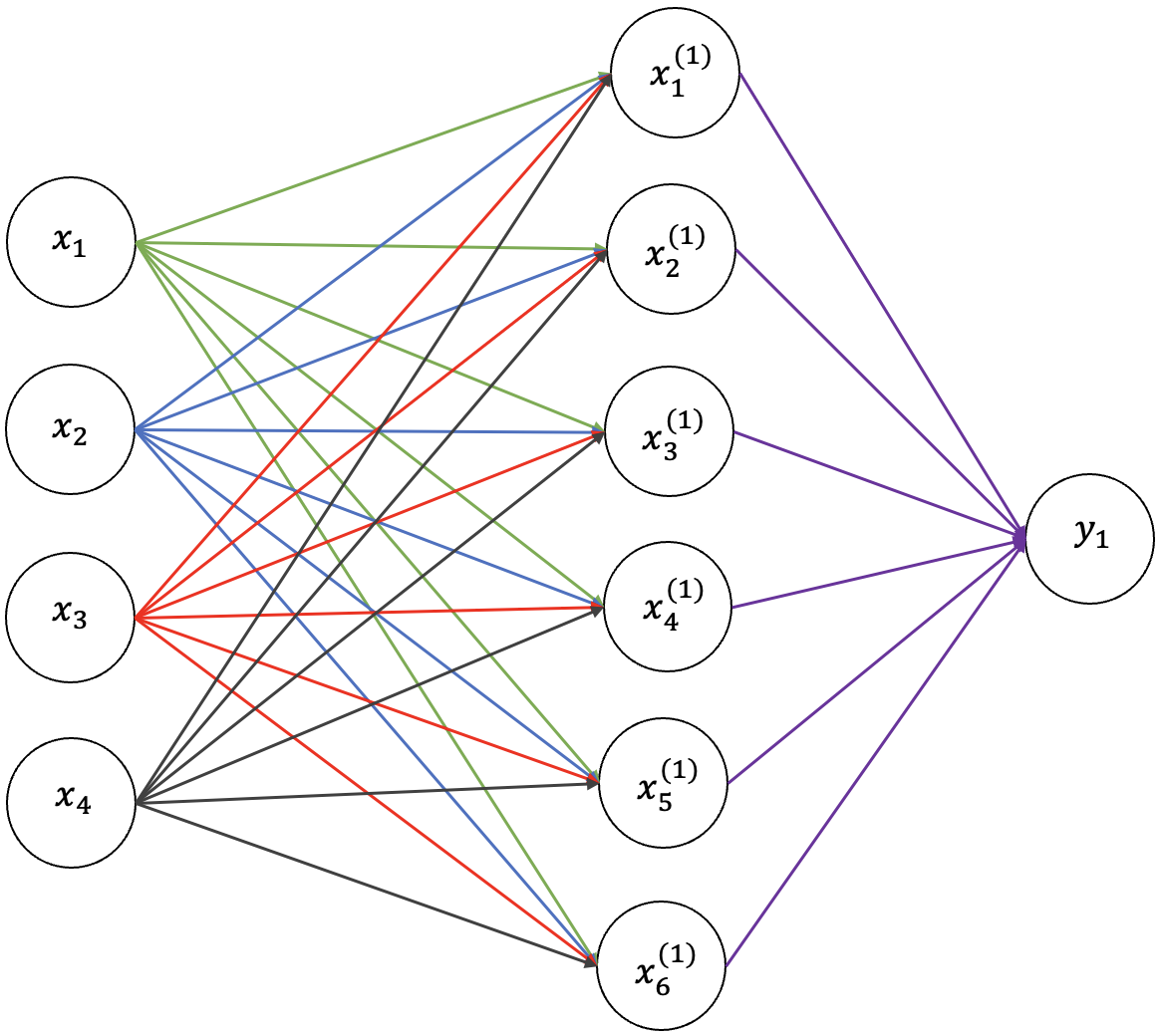}
    \caption{Basic NN structure. The colored lines represent the connections between the neurons from one layer to another. These connections include the weights, biases, and activation functions that transform the data in one neuron to become part of the input of the neurons in the next layer.}
    \label{NNstructure}
\end{figure}

In our case, the input layer possesses $n$ neurons, where $n$ is the number of considered JWST filters. Each of the four variables (redshift, stellar mass, halo mass or mass-weighted age) is predicted by its own NN, so each neuron in the input layer receives an $N \times 1$ vector as input, where $N$ is the number of magnitudes (i.e., number of different samples) for a given filter. Thus, the overall input to the NN is an $N \times n$ matrix. The output layer has one neuron containing an $N \times 1$ vector, which corresponds to the outputs of one of the variables. We consider multi-layered NNs with different numbers of layers. The number of layers $\ell$ of an NN is defined here as the number of hidden layers plus one. Hidden layers in multi-layered NNs can have an arbitrary number of neurons. For a layer $k$, the input matrix $X_k$ is $N \times n_k$, where $n_k$ is the number of neurons in the layer. This input matrix can be related to the input $X_{k+1}$ of the next layer by a $n_k \times n_{k+1}$ weight matrix $W_k$, and a $N \times n_{k+1}$ bias matrix $B_k$, acted on by an activation function $f_k$:

\begin{equation}
    \label{eq:multilayNN}
    X_{k+1} = f_k \left(X_k W_k+B_k \right),
\end{equation}
where 
\begin{equation}
    \label{eq:multilayNN2}
    X_{k} = 
    \begin{cases}
        N \times n \ \rm{input \ matrix} & \text{if } k = 1\\
        f_{k-1}\left(X_{k-1}W_{k-1}+B_{k-1} \right) & \text{if } k > 1
    \end{cases}
\end{equation}
and $k=1$ is the input layer, $k=\ell+1$ is the output layer, and $\rm{y}=X_{\ell+1}$ is the final output.  

For our NNs, we use $6$ neurons for the first hidden layer and $5$ neurons for the rest of the hidden layers. The chosen activation function is applied to all layers except for the final layer, which uses a linear activation function. The architecture of the two-layer NN with four filters is shown in Fig. \ref{NNstructure}. 

The NN is initialized with random weights and biases using the default methods in \textsc{tensorflow}. For the weights, the default is the Glorot Uniform distribution \citep{pmlr-v9-glorot10a}, which contains values in the interval $[-1,1]$. The biases are initially set to zero. The output of the NN is compared with the “true” output from the simulation (the “ground truth”), and the difference between the two is the loss. The loss is quantified by a function, such as the mean squared error (MSE) or the mean absolute error (MAE):
\begin{equation}
    \label{eq:MSE_loss}
    \rm{MSE} = \frac{1}{N}\sum_{i=1}^{N}\left(y_{true}-y_{pred} \right)_i^{2},
\end{equation}
\begin{equation}
    \label{eq:MAE_loss}
    \rm{MAE} = \frac{1}{N}\sum_{i=1}^{N}\left|y_{true}-y_{pred} \right|_i,
\end{equation}
where $\rm{y}_{\rm{true}}$ is the true output from the simulation and $\rm{y}_{\rm{pred}}$ is the output of the NN. To minimize the loss, the NN undergoes a process called backpropagation, in which the weights are adjusted starting from the outermost layer and ending with the input layer \citep{Rumelhart_1987}. A key parameter that determines the magnitude of the adjustment is the learning rate, a value that scales the adjustment of the weights. A carefully selected learning rate is crucial because a learning rate that is too large might lead to a divergent solution, and a learning rate that is too small takes unnecessarily long and might get stuck inside a local minimum \citep{Werbos_1974,Schmidhuber_1989}. {\sc Tensorflow} allows the user to specify loss function, learning rate, activation function, number of layers, and much more. 

\subsection{Parameter Selection}
\label{params}
\par Given seven JWST filters (F150W, F200W, F277W, F356W, F444W, F162M, and F182M), all possible combinations of three, four, five, six, and seven filters are examined in the NNs. The other wide JWST filters are not included in this analysis because the magnitudes of the simulated galaxies are not detectable with those filters. The two medium filters are included to pinpoint the Lyman-alpha break with more precision, improving redshift prediction. Accurate redshift predictions play an important role in the recovery of the other variables. Due to the relationship between distance, light, and mass, predicting redshift is helpful for predicting stellar mass and halo mass. Distance is also related to age, so redshift is a good indicator of mass-weighted age. 

\begin{table}[ht]
\centering
\caption{Activation functions used for each variable}
\begin{tabular}{p{1.5cm}p{1.5cm}p{1.5cm}p{1.5cm}} 
 $z$ & $M_{*}$ & $M_h$ & $t$ \\
 \hline
 elu & elu & relu & elu  \\
 gelu & gelu & linear & gelu \\
 tanh & tanh & tanh & tanh \\ 
 swish & selu & swish & selu \\
 softplus & ... & softplus & softplus \\
 softsign & ... & ... & softsign  \\
\hline
\multicolumn{4}{p{.4\textwidth}}{\emph{Notes:} For each variable, a NN is run for each activation function and each $\ell = 2, 10, 15, \ \rm{or} \ 20$ layers. The less well-known activation functions are defined in the Appendix (section \ref{actv}).}
\end{tabular}
\label{input-parameters}
\end{table}

To optimize the performance of the NNs, we investigate a parameter space consisting of different activation functions, numbers of layers, and filter combinations. These parameters are fixed in each NN but vary as we analyze the performance of multiple NNs. The activation function applies to all layers of the NNs except for the final layer, which has a linear activation function. The activation functions used in the NNs are displayed in Table \ref{input-parameters}, for redshift $(z)$, stellar mass $(M_{*})$, halo mass $(M_h)$, and mass-weighted age $(t)$ (note that stellar and halo mass are in log scale). The activation functions were selected based on an initial performance test of $13$ activation functions available on {\sc Tensorflow}. For each activation function and each variable, we run four NNs, each with a different number of layers: $2$, $10$, $15$, or $20$. 

We choose a learning rate of 0.05 because preliminary tests indicated that learning rates around this value yielded the most accurate predictions, relative to others we tested in the interval $[0.01,0.5]$. These tests also produced better predictions with the MAE loss function and the Stochastic Gradient Descent (SGD) optimizer in {\sc Tensorflow}, so we adopt these parameters in our NNs. We use $85\%$ of our data for training and $15\%$ for testing. However, we use slightly different parameters for halo mass. Initial tests on the halo mass data revealed that those parameters lead to signs of overfitting. We find that the NNs better learn the features of halo mass when the training ratio is $50\%$ and the Adam optimizer is used. The greater amount of test data for halo mass contributes to reducing the spread in the test sample, which is helpful due to the inherent difficulty in predicting halo masses in our chosen mass range ($7.2 \lesssim M_h \lesssim 8.4$). To eliminate any possible relationship between the training and testing samples, we split the data by root halo. This ensures that the data is separated not only by halo, but also by the halos that merged to form the halo. The NNs train on 500 epochs to ensure that there are enough opportunities for the weights and biases to be adjusted while minimizing computation time. To account for statistical variance, we rerun a sample of the NNs with different random seeds used to split the training and validation datasets. This allows us to not only evaluate the statistical uncertainties of our approach, but to also investigate the NN's ability to train on different galaxies in our data.

Prior to entering the data into the NNs, we normalize the training and testing data because many activation functions are most sensitive in the interval $[-1,1]$. The normalization is done by subtracting the mean of the training set and then dividing by the standard deviation of the training set. Once the NN has trained on the data and made predictions, the output data is renormalized to physical values. 

The performance of the NNs in predicting redshift is evaluated through various error metrics (see Eqs. \ref{eq:NRMS_z}-\ref{eq:NMAD}). Normalized Root Mean Squared (NRMS), normalized bias $\langle \Delta z \rangle$, and Normalized Median Absolute Deviation (NMAD) are commonly used in the literature \citep{Bisigello12016,Gomes_2017,Wilson_2020,Zhou_2022b}. These metrics are defined by 
\begin{equation}
    \label{eq:NRMS_z}
    \rm{NRMS} = \sqrt{\frac{1}{N}\sum_{i=1}^{N}\left(\Delta z\right)_i^{2}},
\end{equation}
\begin{equation}
    \label{eq:bias_z}
    \langle \Delta z \rangle = \frac{1}{N}\sum_{i=1}^{N} (\Delta z)_i,
\end{equation}
and
\begin{equation}
    \label{eq:NMAD}
    \rm{NMAD} = 1.4826 \times \rm{median} \left(\left|\frac{\delta z - \rm{median}(\delta z)}{1+z_{\rm{sim}}} \right|\right),
\end{equation}
where $\Delta z = (z_{\rm{sim}}-z_{\rm{pred}})/(1+z_{\rm{sim}})$ and $\delta z = z_{\rm{sim}}-z_{\rm{pred}}$. Another standard metric is the percentage of outliers $\eta$ for which $|\Delta z|>0.05$. Generally, the threshold is 0.15 \citep{Wilson_2020,Zhou_2022b}, but in our case there were too few outliers under this criterion. We lowered the threshold to $0.05$ to facilitate the comparison of outliers between different configurations of parameters. 

For stellar mass, the accuracy of the predictions is evaluated by Root Mean Squared (RMS), bias $\langle \Delta \rm{log}(M_{*}) \rangle$, and Mean Squared Error (MSE) \citep{Bisigello_2017,Simet_2019,Kamdar_2016}: 
\begin{equation}
    \label{eq:RMS}
    \rm{RMS} = \sqrt{\frac{1}{N}\sum_{i=1}^{N}\left(\Delta \rm{log}(M_{*})\right)_i^{2}},
\end{equation}
\begin{equation}
    \label{eq:bias_stellar}
    \langle \Delta \rm{log}(M_{*}) \rangle = \frac{1}{N}\sum_{i=1}^{N} \Delta \rm{log}(M_{*})_i,
\end{equation}
\begin{equation}
    \label{eq:MSE_stellar}
    \rm{MSE} = \rm{RMS}^2,
\end{equation}
where $\Delta \rm{log}(M_{*}) = \rm{log}(M_{*,\rm{sim}})-\rm{log}(M_{*,\rm{pred}})$.

Halo mass predictions are evaluated using MSE and $R^2$ \citep{Zhou2022}: 
\begin{equation}
    \label{eq:MSE_halo}
    \rm{MSE} = \frac{1}{N}\sum_{i=1}^{N}\left(\Delta \rm{log}(M_{h})\right)_i^{2},
\end{equation}
\begin{equation}
    \label{eq:Rsq_halo}
    R^2 = 1-\frac{\sum_{i=1}^{n}\Delta \rm{log}(M_{h})_i}{\sum_{i=1}^{n}\rm{log}(M_{h,\rm{sim}})_i-\langle \rm{log}(M_{h,\rm{sim}})_i \rangle},
\end{equation}
where $\Delta \rm{log}(M_{h}) = \rm{log}(M_{h,\rm{sim}})-\rm{log}(M_{h,\rm{pred}})$.

Predictions of mass-weighted age are evaluated by RMS:
\begin{equation}
    \label{eq:NRMS_mwa}
    \rm{RMS} = \sqrt{\frac{1}{N}\sum_{i=1}^{N}(\Delta t)_i^{2}},
\end{equation}
where $\Delta t = t_{\rm{sim}}-t_{\rm{pred}}$.

\subsection{Observational Constraints} 
\label{constraints}
\par We calculate the AB magnitudes for the merger histories of approximately 180 galaxies from the simulation. There are $237$ time steps in the simulation, but not all halos exist for the entire time interval, so the original data set is composed of $N = 10,426$ samples. This includes data for all halos and every timestep at which they exist and present stellar activity. For the NNs to train on more data under the magnitude limit, we add additional samples to the dataset consisting of the original samples shifted forward in time by $50 \ \rm{Myr}$, $100 \ \rm{Myr}$, and $150 \ \rm{Myr}$. The full dataset then contains $41,704$ samples. We also test the performance of our NN on real objects at $10 < z < 12$ (see section \ref{testcase}), which is additional motivation for using shifted data and obtaining more samples at lower redshifts. We find that training on the shifted data drastically improves predictions of all the variables.

In the analysis of redshift only, we subtract the filter magnitudes in each sample by the lowest magnitude in the sample, so that the NNs learn how to determine the redshift of objects based solely on their spectral shape and regardless of their inherent brightness. Our initial tests revealed that redshift and inherent brightness are less correlated than redshift and the shape of the spectra, so we conclude that intrinsic brightness may be a form of noise that can be helpful to remove. We find that this subtraction significantly increases the accuracy of redshift predictions. 

\par We apply cuts to the data to account for observational constraints and interests:
\par 1) Only use samples corresponding to $z>10$.
\par 2) Only include samples where at least two filters have a magnitude brighter than the magnitude limit, to reduce redundant information. Then set all magnitudes fainter than the magnitude limit equal to the magnitude limit to mimic real observations. 
\par 3) Only include filter combinations where the maximum redshift is at least 15, so we can study galaxies in our full target range $11.5 < z < 15$. Also, only include combinations where the maximum redshift of both the training and test set is at least 14.25. This reduces major differences in the values between the training and test sets. 

After these cuts are applied, the number of samples in our data set drops to approximately $20,000 \pm 3,000$, depending on the filter combination. The NNs train on this data, but ultimately our goal is to assess the NN's ability to make predictions with the original data. Thus, we use the original, unshifted data to obtain the test set. In other words, none of the test galaxies interact with the augmented dataset in any way.  Since we validate our results on the unshifted data only, the augmented training data introduce no physical bias. After cuts are applied, the final size of the unshifted dataset is roughly $6,000 \pm 1,000$ samples. Since the root halos are the same for both the shifted and unshifted data, they are randomly split between training and testing in the same way. 

We set the magnitude limit at $\rm{ABmag} = 36$, based on the approximate sensitivity limit of $31$ at a signal-to-noise ratio (SNR) of $10$ and exposure time of $100$ hours for the JWST filter F200W, according to the Exposure Time Calculator (ETC) v2.0 \citep{Pickering_2016}. The magnitude difference is accounted for by the ability to detect sources up to 5 magnitudes fainter with gravitational lensing \citep{Richard_2014,Johnson_2014}. We must also consider JWST's maximum visit time of 100 hours, according to the Astronomer's Proposal Tool (APT) v2022.3.1 \citep{Roman_2004}. Given that each filter has its own sensitivity, the SNR achieved by each filter in $100$ hours to detect $\rm{ABmag} = 31$ differs substantially, as shown in Table \ref{vis}. In addition, the exposure times required to achieve SNR $\geq 5$ varies considerably for each filter. This table also records what percentage of the magnitudes are brighter than 36 for each filter, taking into account the gravitational lensing effect. 

\begin{table}[ht]
\centering
\caption{SNRs, visibility percentages, and exposure times for each of the JWST filters considered in this work} 
\begin{tabular}{cccc} 
 \multirow{2}{*}{Filter} & \multirow{2}{*}{SNR (100 h)} & \multirow{2}{*}{Visibility \%} & Exposure time (h) \\ & & & to SNR = 5 \\
 \hline
 F150W & 13.5 & 41.7 & 10.7\\
 \hline
 F200W & 13.5 & 67.2 & 10.8\\
 \hline
 F277W & 9.0 & 63.4 & 24.0\\
 \hline
 F356W & 7.0 & 56.8 & 42.5\\
 \hline
 F444W & 3.1 & 51.3 & 663.5\\
 \hline
 F162M & 9.5 & 55.0 & 25.2\\
 \hline
 F182M & 10.4 & 66.1 & 20.7\\
 \hline
\multicolumn{4}{p{.45\textwidth}}{\emph{Notes:} The SNR shown for each filter is for 100 hours of exposure time at $\rm{ABmag} = 31$. The visibility percentage corresponds to the fraction of filter magnitudes in our shifted data set that are less than 36 (after the first cut is applied and before the second). The exposure time is the number of hours required to reach $\rm{ABmag} = 31$ with SNR = 5. The SNRs and exposure times are calculated based on ETC v2.0, with an aperture radius of $0.04$" and a $0.58\rm{"}-0.7\rm{"}$ background sky annulus. The background is assumed to be the benchmark background model (RA=$17$h $26$m $44$s, Dec=$-73^{\circ}19$'$56$" on May 5, 2022) \citep{Rigby2022}.}
\end{tabular}
\label{vis} 
\end{table}

In order to take into account the observational feasibility of each filter combination, we present results for an unconstrained case, an observationally feasible case, and an observationally ideal case. The unconstrained case considers every mathematically possible combination for three to seven filters. For the observationally feasible case, we consider combinations with four or six filters and select the combinations that have an equal number of short wavelength (SW) filters (F150W, F200W, F162M, F182M) and long wavelength (LW) filters (F277W, F356W, F444W). These SW and LW filter pairs do not necessarily need to have similar sensitivities because the same observing time is used for all filters in this scenario (100 hours, the maximum).  

For the observationally ideal case, we determine the exposure time required to get $\rm{SNR} = 5$ at $\rm{ABmag} = 31$ for each filter (see Table \ref{vis}). We consider combinations with four filters and select the combinations that have not only an equal number of SW and LW filters, but similar sensitivities between the filter pairs. This allows us to pair the SW and LW filters by exposure time, to take advantage of the parallel observing feature between SW and LW filters. For this case, we exclude F444W due to its extremely long exposure time of 663.5 hours. 

\section{Results}
\subsection{Best Unconstrained Results}
\label{theo_results}

\begin{figure*}[ht]
    \centering
    \includegraphics[width=14.5cm]{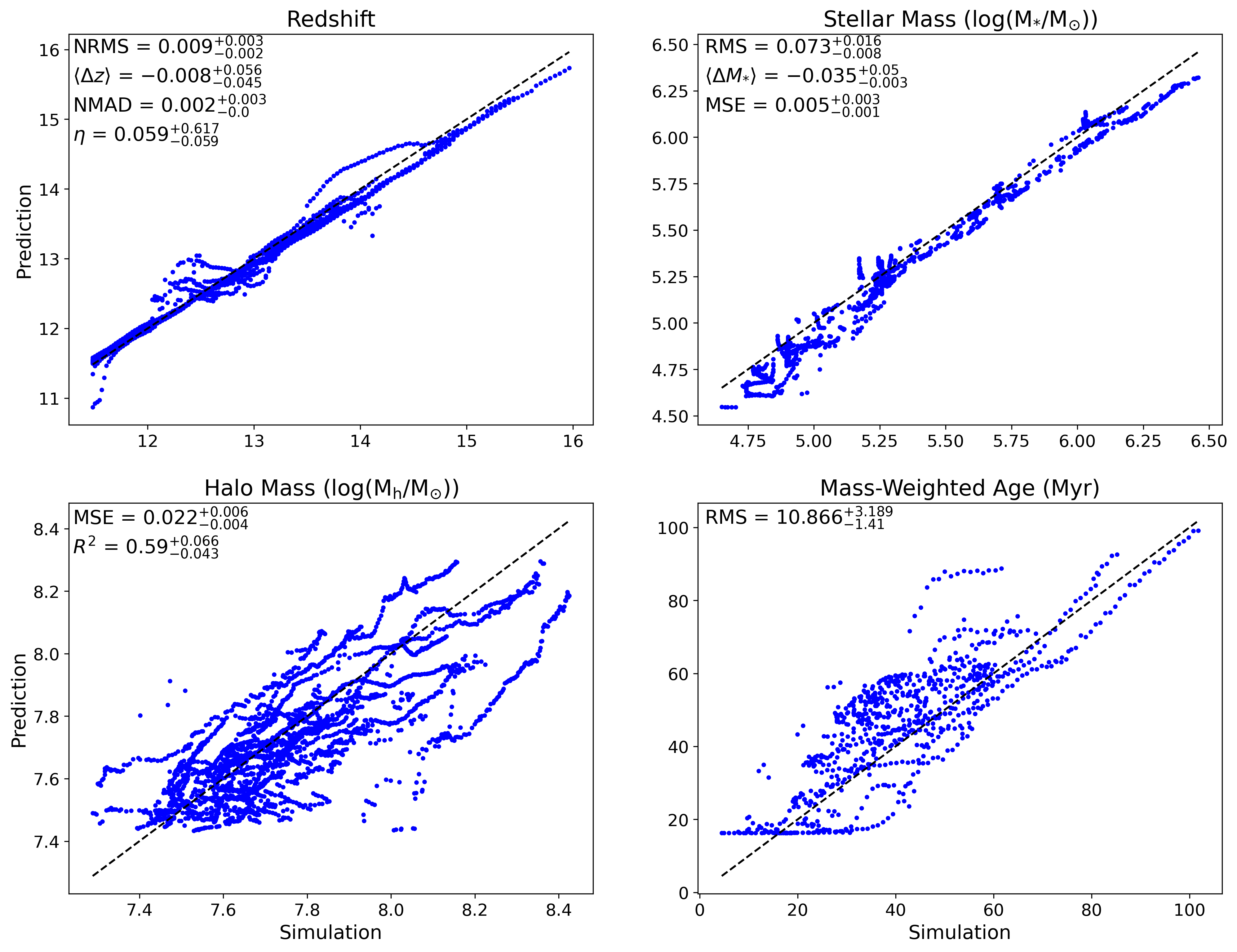}
    \caption{Prediction vs. simulation of the test data from the best NN for each variable, in the unconstrained case. The black dashed line indicates a one-to-one correspondence between the predicted outputs from the NN and true values from the simulation. The top left plot shows that {\sc Gainn} can predict the redshift of galaxies at $11.5 \lesssim z \lesssim 15$ with few errors and outliers. Stellar mass (top right) can also be predicted by {\sc Gainn} with low errors. Predictions for halo mass (bottom left) show considerable scatter, consistent with the loss plot in the appendix indicating the NN's difficulty in learning from the halo mass data. {\sc Gainn} can predict mass-weighted age (bottom right) reasonably well despite some scatter and outliers, as long as the ages are greater than about 18 Myr.}
    \label{theo_res_multilay}
\end{figure*}

\begin{table*}[ht]
\centering
\caption{Input parameters of the best NN for each variable and each case}
\begin{tabular}{ccccccccccc}
\multirow{2}{*}{Variable} & \multirow{2}{*}{$f$} & \multirow{2}{*}{Layers} & \multicolumn{6}{c}{\multirow{2}{*}{Filters}} & Samples & Samples \\ & & & & & & & & & (shifted) & (unshifted) \\
\hline
\multicolumn{11}{c}{Unconstrained} \\
\hline
z & elu & 10 & F150W & F200W & F277W & F162M & F182M & ... & 23019 & 6714 \\ 
\hline
$M_{*}$ & elu & 15 & F150W & F200W & F277W & F356W & F162M & ... & 22656 & 6704 \\ 
\hline
$M_h$ & swish & 2 & F150W & F356W & F444W & F162M & ... & ... & 19899 & 5707 \\ 
\hline
t & softplus & 10 & F200W & F277W & F356W & F444W & F162M & ... & 22605 & 6704 \\ 
\hline

\multicolumn{11}{c}{Feasible} \\
\hline
z & elu & 10 & F200W & F277W & F356W & F444W & F162M & F182M & 23242 & 6940 \\ 
\hline
$M_{*}$ & gelu & 10 & F150W & F200W & F277W & F356W & F444W & F162M & 22656 & 6704 \\ 
\hline
$M_h$ & swish & 2 & F150W & F356W & F444W & F162M & ... & ... & 19899 & 5707 \\ 
\hline
t & gelu & 2 & F200W & F277W & F356W & F444W & F162M & F182M & 23242 & 6940 \\ 
\hline

\multicolumn{11}{c}{Ideal} \\
\hline
z & swish & 10 & F277W & F356W & F162M & F182M & ... & ... & 22670 & 6673 \\ 
\hline
$M_{*}$ & elu & 10 & F277W & F356W & F162M & F182M & ... & ... & 22670 & 6673 \\ 
\hline
$M_h$ & linear & 2 & F150W & F277W & F356W & F162M & ... & ... & 21325 & 6319 \\ 
\hline
t & softplus & 10 & F200W & F277W & F356W & F162M & ... & ... & 22605 & 6704 \\ 
\hline
\multicolumn{11}{p{.9\textwidth}}{\emph{Notes:} This table reports the set of hyper-parameters (activation function, number of layers, and filter combination) that leads to the lowest error in predicting each variable in each case. The last two columns indicate the final number of samples in the shifted and unshifted datasets for each filter combination, after all cuts have been applied. In the observationally feasible and ideal cases, a single filter combination provides the most accurate prediction of redshift and one other variable (mass-weighted age for the feasible case and stellar mass for the ideal case). Thus, more than one variable can be predicted at high accuracy with a single JWST filter combination. This combination depends on the availability of JWST resources.}
\end{tabular}
\label{all_input_params}
\end{table*}

The best unconstrained results are the best predictions made by a NN when all possible combinations of filters, from three to seven filters, are accounted for. In this scenario, a total of ninety-four combinations are considered, excluding the five combinations removed by the third cut. Considering each possible activation function and number of layers in Table \ref{input-parameters}, this amounts to 2,256 configurations for redshift and mass-weighted age, 1,880 configurations for halo mass, and 1,504 configurations for stellar mass. The configurations are sorted by NRMS for redshift, by MSE for halo mass, and by RMS for stellar mass and mass-weighted age. We then take the top 20 filter configurations for for each set (giving a total of 100 configurations) and rerun each configuration with 51 different seeds. For each configuration, we take the median across all 51 seeds (including the original) of the metric used to sort the configurations (NRMS, RMS, or MSE). The medians are then sorted and the configuration with the lowest median is taken to be the best for a given variable, with an uncertainty described by the 15th and 85th percentile of the sorting metric across all seeds. The results presented here are those corresponding to the best configuration. For some configurations, however, there is no data for one of the seeds due to the third cut. Then these configurations have 50 seeds, and we take the median to be the higher of the two mid-values.

Due to the visibility constraint imposed on the data (at least two filters need to display a magnitude brighter than the magnitude limit), there are fewer data points corresponding to farther, older, high-redshift galaxies. Therefore, we expect it to be difficult for the NNs to predict higher redshifts and mass-weighted ages as a result. We also anticipate difficulty in detecting properties of dimmer, low-mass galaxies, but more accurate predictions for galaxies with greater halo masses and total stellar masses. As we explain below, these features are generally what we observe in our results. 

\begin{figure*}[ht]
    \centering
    \includegraphics[width=14.5cm]{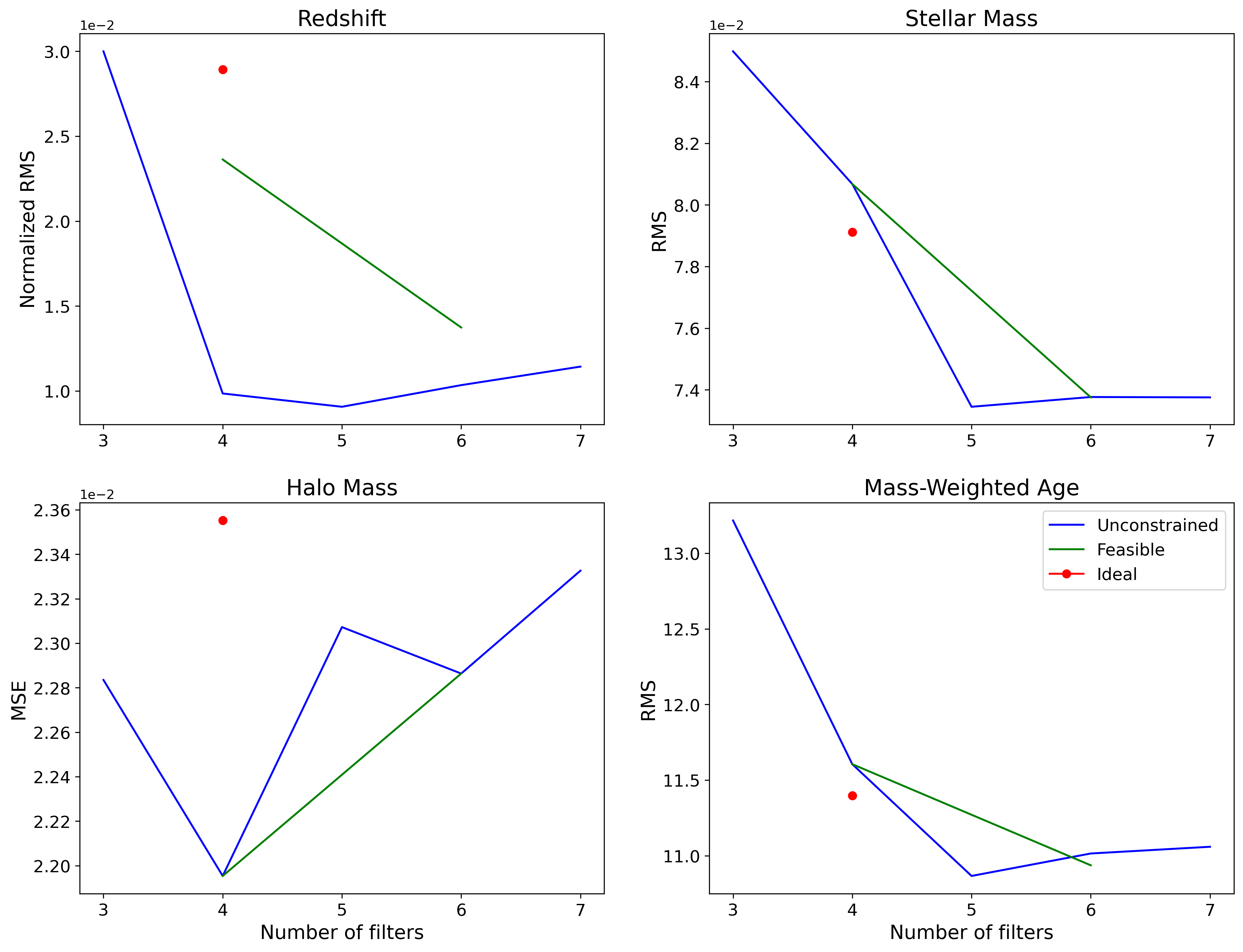}
    \caption{Error vs. number of filters for all variables in all cases, given the best results for each number of filters. In the unconstrained case (blue), the error has a global minimum at five filters for every variable other than halo mass. Also, with the exception of halo mass, the error for all other variables decreases with increasing number of filters in the observationally feasible case (green). Since we only consider combinations with four filters in the observationally ideal case (red), its error is represented by a single point for each variable.}
    \label{error_nfilt}
\end{figure*}

Fig.  \ref{theo_res_multilay} shows the relationship between the output of the NNs (the prediction) and the results from the simulation (the ground truth), for the best NNs of each variable. The closer the data is to modelling the black dashed line, the better the NN is at predicting the values of a given variable. Plots showing the loss as a function of epochs are included in the Appendix (section \ref{loss}). Overall, the losses roughly converge for each of the variables, though for halo mass the loss is extremely oscillatory and demonstrates difficulty in converging. This is an indication of the inherent difficulty in using filter fluxes to predict halo masses in a range as low as ours.

The best NN's predictions of redshift closely match the true values, as displayed in Fig. \ref{theo_res_multilay}. The NRMS of this NN ($0.009_{-0.002}^{+0.003}$) is comparable to that of \citep{Bisigello12016}, \citep{Gomes_2017}, and \citep{Wilson_2020}, with NRMS $< 0.01$, $\sim 0.03$, and $\sim 0.1$ respectively. Our redshift range covers several filter transitions, so there is some variation when transitioning between filters. These
errors occur because there is a slight gap in the throughputs between the different filters, and thus there is no data in a small region of the wavelength space. The NN presents more scatter and a greater number of outliers at $z \sim 12-13$ and $z \sim 14$ as a result of the change in filters at these redshifts. The filter transition at z=12.5 is between F150W and F200W, and the transition at z=14 is F200W to F277W. The slight bias of the NN's redshifts from the true values at $z \gtrsim 14$ occurs due to the low availability of training data at $z \gtrsim 14$ for that particular filter combination and seed. Only $\sim 5\%$ of the training data for redshift lies in that range. Other than the filter transitions and the small deviation at $z \gtrsim 14$, the relationship between the predicted and simulated redshifts is very tightly correlated.

As expected, the predictions of stellar mass are generally more accurate at higher values. An exception is the bias observed at $M_{*} \gtrsim 6.2$, where the NN underestimates the values of stellar mass. This could be explained in part by limited training data ($\sim 4\%$) in that range. There is also more bias at $M_{*} \lesssim 5.25$, but this is not surprising given the greater difficulty in predicting lower stellar masses. The result for halo mass requires a different interpretation. As stated previously, the loss plot of halo mass indicates that the NN could not learn enough about this variable with the information it was given. Therefore, the relationship between the true and predicted halo masses shown in Fig. \ref{theo_res_multilay} is not a much better fit than what a random guess could provide (which is what the NNs begin with). Nonetheless, the plot does show some correlation between the simulated and predicted halo masses from the NN. In fact, the best NNs predicting stellar mass and halo mass perform well relative to the results in the literature. The RMS for stellar mass ($0.073_{-0.008}^{+0.017}$) is comparable to that of \citep{Bisigello_2017}, which presents RMS no lower than $0.04$. The MSE for halo mass ($0.022_{-0.004}^{+0.006}$) is comparable to $0.01$, presented by \citep{Zhou2022}. Both references use mass ranges higher than ours. Since higher stellar and halo masses correspond to brighter galaxies and therefore more information to train on, the comparison between our results is conservative.
Results for mass-weighted age have more scatter and outliers than redshift and stellar mass. Surprisingly, predictions of higher mass-weighted ages do not appear to be worse than lower mass-weighted ages, contradicting our expectations. In fact, the flat line at predicted mass-weighted ages of $\sim 18$ Myr suggests that the NN cannot recognize when a galaxy is younger than $\sim 18$ Myr. This is likely not due to low availability of training data, given that $\sim 22\%$ of the training data lies in the range $t \lesssim 18$ Myr. This is likely due instead to the high variability in young star cluster spectra. Therefore, this NN should only be used to predict mass weighted ages higher than $\sim 18$ Myr.

\begin{figure*}[ht]
    \centering
    \includegraphics[width=15.5cm]{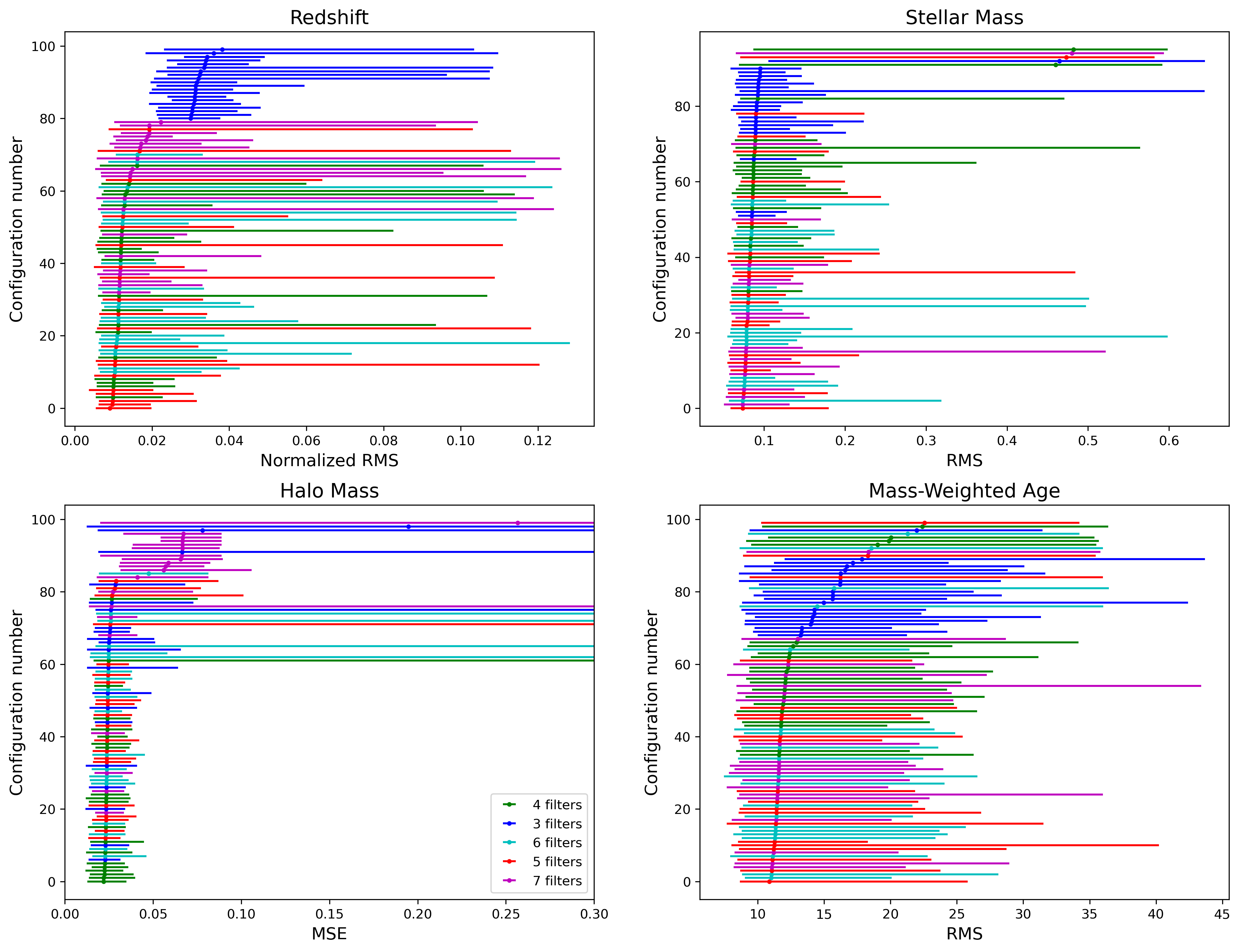}
    \caption{Error spread across seeds of all configurations. Each line represents a configuration and ranges from the minimum to maximum error. The points correspond to the median errors of the configurations. Thus, the lines show the full range of errors among the seeds for each configuration.}
    \label{error_plot}
\end{figure*}

Table \ref{all_input_params} reports the input parameters that led to the best results for each variable. To provide more information on the other configurations, we also discuss which filter appears most frequently in the 100 selected configurations for each variable. In the case of redshift, the most frequently occurring filter is F182M, which agrees with the appearance of this filter in the NN with the best redshift predictions. In fact, it is crucial to note that F182M appeared in every single configuration of the selected 100 for redshift, making it an indispensable filter for redshift prediction. F200W is the most common filter in the selected configurations for mass-weighted age. The corresponding filter for stellar mass is F150W. For halo mass, the most frequently appearing filter is F444W. Consistently, the most common filter in the selected configurations for each variable is also one of the filters in the best configuration for each variable. Therefore, the most predictive filters across all variables and configurations are F182M, F200W, F150W, and F444W. However, this four-filter combination is not very practical when making observation proposals with JWST. The combination has only one LW filter and three SW filters, removing the opportunity to take advantage of JWST's parallel observing capabilities. This issue is accounted for in sections \ref{interm_results} and \ref{obs_results}, in which a more appropriate balance between SW and LW filters is considered in addition to theoretical accuracy. 

The relationship between the error and the number of filters used as inputs in the NN is illustrated in Fig. \ref{error_nfilt}. For all variables except for halo mass, there is a tendency for the error to decrease between three and five filters, and then increase again when more filters are added. It is expected that the error would generally decrease with an increasing number of filters, due to the additional information that more filters would provide. However, adding more filters also adds more noise. This occurs here because a greater number of filters implies a greater probability that some data points will have most of the filter magnitudes equal to the magnitude limit, given that only two of the filter magnitudes need to be brighter than the magnitude limit for the data point to pass the condition of the second cut. The neural network only sees the magnitude limit as a bound, so magnitudes at the magnitude limit do not provide any useful information. In other words, the NN is only training on the magnitudes below the magnitude limit, which may be distinguishable from one another.

The result of the competing factors that adding filters provides more information but also more noise is the global minimum observed at five filters. The dominant factor is the former, given that there is a significant drop in error between three and five filters compared to the marginal increase between five and seven filters. This is an indication that adding more information can be extremely beneficial to the NN's ability to make predictions and might outweigh any potential addition of noise.

Halo mass does not exhibit this behavior, however. The minimum occurs at four filters and then the error increases significantly at five filters. This could be a consequence of the NN's difficulty in training on the halo mass data. Since the results for halo mass are not much better than a random guess, it is more likely that they could deviate from expected behavior.

Fig. \ref{error_plot} displays plots illustrating the spread of error across seeds for all configurations. The variation among the different seeds demonstrates how well each NN trains on different datasets. The configurations with shorter lines correspond to NNs that learn well regardless of the provided datasets, whereas longer lines indicate that the NNs have trouble with a large variety of input data. Since each seed offers a unique set of galaxies for the NNs to train on, the length of the line is a measure of how well the NNs learn from different galaxies. The location of the median point is a measure of how well a given NN performs relative to others overall. For most variables, the medians are very close to each other up to a point, but then tend to deviate when more configurations with three filters are considered. This is consistent with our earlier discussion regarding the benefit to having more filters to provide input data. The wide variety of line sizes suggests considerable variation in how the NNs handle various datasets. Some NNs predicting halo mass could not make accurate predictions with certain seeds, which led to very large errors not shown in the plot.

\subsection{Best Observationally Feasible Results}
\label{interm_results}

 In the observationally feasible scenario, we add constraints that take into account JWST resources and observing capabilities. We continue adopting the maximum exposure time of 100 hours for each filter, but we only consider combinations with an equal number of SW and LW filters. This narrows our results to configurations with four or six filters that have two or three of each filter type. The resulting number of filter combinations is twenty-two, leading to 528, 440, and 352 configurations for redshift and mass-weighted age, halo mass, and stellar mass respectively. As in the unconstrained case, we rerun the top 20 configurations for each number of filters with 50 new seeds and take the median of the sorting metric from each configuration. In this section, we present the results for the configuration with the lowest median for each variable in the feasible case.

Table \ref{metrics_interm_obs} documents the metrics of the best NNs in all cases. It is possible for the same NN to yield the best results in both the unconstrained and observationally feasible cases. This can occur if the best filter combination in the unconstrained case also satisfies the conditions of the feasible case. All input parameters and metrics would then be identical in both cases. In fact, the same inputs and metrics are observed between the two cases for halo mass in Tables \ref{all_input_params} and \ref{metrics_interm_obs}, indicating that the same NN produced the best predictions of each variable.

However, in general the permitted configurations of the feasible case do not include the best configuration of the unconstrained case. We therefore expect that the results of the feasible case are usually less accurate than those of the unconstrained case. Most of the metrics are a measure of error in the neural network's predictions, so lower values imply more accurate results (except for $R^2$, which indicates a better fit with higher values). According to Table \ref{metrics_interm_obs}, redshift follows the anticipated trend. The metrics of the feasible case are slightly greater than those of the unconstrained case for redshift (considering the absolute value of the average bias). The RMS of stellar mass and mass-weighted age conforms to the expected trend, but the magnitude of the stellar mass bias is lower in the feasible case and the stellar mass MSE is equal in the two cases. Overall, the differences between the metrics of the feasible and unconstrained cases are marginal. 

As in the unconstrained case, we also consider how often each filter appears in the 40 selected configurations for each variable in the observationally-feasible case. In the selected configurations for redshift and halo mass, the most frequently occurring filter is F444W. For stellar mass and mass-weighted age, F277W is the filter that appears the most. Accordingly, F444W is in the best combination predicting redshift and halo mass, and F277W is in the best combination for recovering stellar mass and mass-weighted age. Therefore, F444W and F277W are the most predictive filters in the feasible case. If F200W, F356W, F162M, and F182M were added to these two filters, the resulting six-filter combination would be the one that provides the best predictions of redshift and mass-weighted age. 

The green line in Fig. \ref{error_nfilt} corresponds to the error with respect to the number of filters for the observationally feasible case. For all variable except halo mass, the error is lower for combinations with six filters than combinations with four filters, following the predicted downward trend. The error in predicting halo mass exhibits an upward trend. As in the unconstrained case, a potential cause for this may be the the NN's difficulty in predicting halo mass.

\subsection{Best Observationally Ideal Results}
\label{obs_results}

\begin{table*}[ht]
\centering
\caption{Final metrics for the best NNs in all cases}
\begin{tabular}{ccccccccccc}
& \multicolumn{4}{c}{} & \multicolumn{3}{c}{} & \multicolumn{2}{c}{} & Mass-weighted\\ 
& \multicolumn{4}{c}{Redshift} & \multicolumn{3}{c}{Stellar Mass} & \multicolumn{2}{c}{Halo Mass} & Age\\ 
\hline
Case & NRMS & $\langle \Delta z \rangle$ & NMAD & $\eta$ & RMS & $\langle \Delta M_{*} \rangle$ & MSE & MSE & $R^2$ & RMS\\
    \hline
        Unconstrained & 0.009 & -0.008 & 0.002 & 0.059 & 0.073 & -0.035 & 0.005 & 0.022 & 0.59 & 10.866 \\ 
       15th/85th perc. & $_{-0.002}^{+0.003}$& $_{-0.045}^{+0.056}$& $_{-0.000}^{+0.003}$& $_{-0.059}^{+0.617}$& $_{-0.008}^{+0.016}$& $_{-0.003}^{+0.050}$& $_{-0.001}^{+0.003}$& $_{-0.004}^{+0.006}$& $_{-0.043}^{+0.066}$& $_{-1.410}^{+3.189}$ \\ 
       \hline
        Feasible & 0.014 & 0.015 & 0.003 & 1.934 & 0.074 & -0.023 & 0.005 & 0.022 & 0.59 & 10.937 \\ 
       15th/85th perc. & $_{-0.004}^{+0.004}$ & $_{-0.076}^{+0.024}$ & $_{-0.000}^{+0.003}$ & $_{-1.291}^{+1.312}$ & $_{-0.010}^{+0.013}$ & $_{-0.002}^{+0.048}$ & $_{-0.001}^{+0.003}$ & $_{-0.004}^{+0.006}$ & $_{-0.044}^{+0.066}$ & $_{-1.516}^{+4.904}$ \\ 
        \hline
        Ideal & 0.029 & -0.02 & 0.013 & 11.475 & 0.079 & 0.003 & 0.006 & 0.024 & 0.627 & 11.398 \\
        15th/85th perc. & $_{-0.004}^{+0.007}$ & $_{-0.210}^{+0.008}$ & $_{-0.007}^{+0.001}$ & $_{-4.732}^{+2.584}$ & $_{-0.011}^{+0.018}$ & $_{-0.030}^{+0.019}$ & $_{-0.001}^{+0.003}$ & $_{-0.004}^{+0.003}$ & $_{-0.066}^{+0.016}$ & $_{-1.453}^{+3.157}$ \\ 
    \hline
\multicolumn{11}{p{.85\textwidth}}{\emph{Notes:} For redshift, the metrics used to evaluate the performance of the NN are NRMS, $\langle \Delta z \rangle$, NMAD, and $\eta$. Stellar mass predictions are evaluated by RMS, $\langle \Delta M_{*} \rangle$, and MSE. The corresponding metrics for halo mass are MSE and $R^2$. Finally, mass-weighted age predictions are assessed by the RMS metric. The results from the unconstrained case are almost always better, but not significantly. Thus, the observationally feasible and ideal cases can be used to create JWST proposals with filter combinations that lead to accurate predictions while preventing exhaustive use of JWST resources.}
\end{tabular}

\label{metrics_interm_obs}
\end{table*}

The observationally ideal results are another subset of the unconstrained results. Through this scenario, we propose alternative observing strategies that minimize exposure times and avoid exhaustive use of JWST resources. Here we include combinations of four filters that contain pairs of SW and LW filters with similar exposure times. According to Table \ref{vis}, there are five pairs of an SW and LW filter with a difference in exposure time less than 20 hours: F277W-F150W, F277W-F200W, F277W-F162M, F277W-F182M, and F356W-F162M. Thus, we classify these as base pairs. In the four-filter case, there are three combinations with two base pairs of distinct filters: F277W-F182M-F356W-F162M, F277W-F200W-F356W-F162M, and F277W-F150W-F356W-F162M. Each filter combination is run through the NN with a set of other changing parameters (activation function and number of layers), so there are still many configurations considered in this scenario (72 for redshift and mass-weighted age, 60 for halo mass, and 48 for stellar mass). Once again, we rerun the top 20 configurations for each number of filters and take the median across the 51 seeds of each configuration. We present the results of the configuration with the lowest median for each variable in the ideal case.  

It is expected that the best observationally ideal results will be worse than the best unconstrained results, due to the same explanation provided in the comparison of the best feasible and unconstrained results. Less stringently, the best ideal results are expected to be worse than the best feasible results, because fewer configurations are considered in the ideal case. This reduces the probability that one of the configurations will yield better results than in the observationally feasible case. While these trends are generally observed in Table \ref{metrics_interm_obs}, there are two exceptions: the average bias of stellar mass and the $R^2$ of halo mass. The metrics of all variables except for redshift have overlapping uncertainties among all cases. Overall, the discrepancies between the metrics of the three cases are not significant, allowing for more practical observing strategies that retain much of the accuracy provided by the unconstrained best results.

Once again, we consider the most frequently appearing filters in the 20 selected configurations for each variable. The top filters for all variables are F277W, F356W, and F162M. Hence, the most predictive filters in the observationally ideal case are F277W, F356W, and F162M. These three filters and F182W would be ideal for a JWST observing proposal because not only are they useful in recovering the variables, but they also form a combination with an equal number of SW and LW filters. This filter combination is also the best for predicting redshift and stellar mass. 

\subsection{Redshift Test Case With Real Objects}
\label{testcase}
The first practical application of {\sc Gainn} was to determine the redshift of the triply lensed dual-clump object MACS0647—JD, originally observed by the Hubble Space Telescope and more recently with JWST. Both primary clumps in MACS0647—JD (A and B), which could potentially be individual galaxies in a merger, had different UV slopes and SED fitting implied both had dissimilar and unique star formation histories. Each projection of the combined object (JD1, JD2, and JD3) had higher signal to noise ratios in JWST wide-band filters F200W, F277W, F356W, and F444W, lower flux in F150W, and signals consistent with no detection in higher wavelength bands. This combination suggests a Lyman break within the F150W filter and therefore a redshift range of $10 < z < 12$. 

As previously shown, our shifted data recovers redshifts from our un-shifted test samples, implying that the redshift-dependent changes to the galaxy formation environment were not impactful enough to skew our predictions. Therefore, we attempted to use our shifted data set to attempt to recover redshifts over the entire range from $10 < z < 12$ and make a prediction for MACS0647—JD with the intention of comparing our results to other predictions and using the multiple projections as a further test of the robustness of our methodology on real observational data. 

In our results, which are more fully presented in  \citet{2022arXiv221014123Y}, the first projection of the combined object (JD1) had dramatically smaller mean absolute errors ($\langle \Delta z \rangle <0.00228$, as calculated using Eq. \ref{eq:bias_z}) in the NN validation set than fits using \textsc{Bagpipes}, \textsc{piXedfit}, \textsc{BEAGLE}, \textsc{Prospector}, and \textsc{EASY}. This is owed in part to our use of fully simulated star formation histories to create our SEDs, which may contain periods of realistically bursty or suppressed star formation that can be challenging to recover or model using other techniques. Our photometric redshift prediction for JD1 ($z=10.4761$) also fell comfortably within the error region for the predictions from each of the other methods ($10.42<z<10.68$), suggesting that our methodology does not measurably bias photometric redshift predictions on real objects, including those slightly outside the redshift range of our simulation. 

Defining $\sigma_z$ as the standard deviation of the predicted redshifts between measurements, each projection of the clumps returned similar redshifts between the clumps ($\sigma_z < 0.063$) and between the projections ($\sigma_z <0.15$), which was consistent with systematic errors in obtaining precise photometeric values for less bright projections and components of MACS0647—JD than JD1. Our $\sigma_z$ values were also smaller than the $\sigma_z$ values for the other methods, further suggesting that remaining measurement differences may be dominated by observational systematics. Errors were also significantly smaller than results from our network when trained on a larger redshift range, so we suggest retraining on a smaller redshift range when looking at individual objects and for cases when precision in photometric redshifts measurements is important, such as when studying possible pre-merger galaxies. 

\section{Summary and Discussion}
\label{disc}
We use NNs to predict the redshift, stellar mass, halo mass, and mass-weighted age of simulated galaxies at $11.5 \lesssim z \lesssim 15$ based on JWST filter magnitudes. To explore the entire parameter space of the NNs, we vary the activation function, number of layers, and filter combination. For each of the top performing configurations of these parameters, we investigate the effect of using different seeds to split the data between training and testing. In theory, we find that F150W, F200W, F182M, and F444W are crucial filters in the study of high-redshift galaxies. In the unconstrained case, {\sc Gainn} can predict redshift, stellar mass, halo mass, and mass-weighted age with errors as low as $0.009_{-0.002}^{+0.003}, 0.073_{-0.008}^{+0.017}, 0.022_{-0.004}^{+0.006},$ and $10.866_{-1.410}^{+3.189}$ (where “errors” refers to NRMS, RMS, MSE, and RMS respectively). 

In addition to reporting the best results considering all possible combinations, we also present results that take into account JWST resources and observing times. The most predictive filters across all variables in the observationally feasible case are F444W and F277W. {\sc Gainn}'s predictions of redshift, stellar mass, halo mass, and mass-weighted age in the feasible case have errors as low as $0.014_{-0.004}^{+0.004}, 0.074_{-0.010}^{+0.013}, 0.022_{-0.004}^{+0.006},$ and $10.937_{-1.516}^{+4.904}$. The accuracy of predictions in the ideal case is very close to the accuracy of the feasible results, indicating that we can interpret the ideal results as a reasonable balance between theory and observational constraints. The most predictive filters across all variables in the ideal case are F277W, F356W, and F162M. {\sc Gainn} achieves errors as low as $0.029_{-0.004}^{+0.007}$, $0.079_{-0.011}^{+0.018}$, $0.024_{-0.004}^{+0.003}$, and $11.398_{-1.453}^{+3.157}$ when recovering redshift, stellar mass, halo mass, and mass-weighted age in the ideal case.

Overall, the most important findings of this work are the following: 

\par (i) The optimal number of filters tends to be around five. Adding more filters is generally helpful for the NN's performance, but at some point adding more filters will add more noise beyond an acceptable limit.

\par (ii) In general, the best results of the observationally feasible and ideal cases are only slightly worse than those of the unconstrained case. This makes it possible to create a JWST proposal that accounts for the span of available resources and is almost as accurate as the unconstrained best-case scenario. 

\par (iii) The most suitable filter combination to use in a JWST observation would be F277W, F356W, F162M, and F182M. The first three filters are the most predictive filters in the ideal case and adding F182M makes this combination the best one for predicting redshift and stellar mass in the ideal case. This combination is also convenient for observing, given that it has an equal number of SW and LW filters with similar sensitivities.

\par (iv) {\sc Gainn} estimated the redshift of MACS0647—JD, an object observed by JWST. The redshift prediction of the object's first projection was $z=10.4761$, with mean absolute errors $\langle \Delta z \rangle < 0.00228$. Estimates from \textsc{Bagpipes}, \textsc{piXedfit}, \textsc{BEAGLE}, \textsc{Prospector}, and \textsc{EASY} did not have errors that low. However, these programs incorporate observational uncertainties into their analyses while {\sc Gainn} does not. We also acknowledge that results may change somewhat with different simulation models.

Relative to results in the literature, the predictions from our best NNs in the unconstrained case are comparable if not more accurate. As for predictions of real observations, our NNs recovered the redshift of a real object observed by JWST with lower errors than other redshift estimation tools. This implies that {\sc Gainn} shows significant promise for redshift recovery in future JWST observations. 

\section*{ACKNOWLEDGMENTS}
LSO was supported in part by the Stanford Physics Department, and by NSF REU funding in initial stages of this work. KSSB was supported by the Porat Postdoctoral Fellowship at Stanford University and the NASA Hubble Postdoctoral Fellowship hosted by Harvard University. TH acknowledges funding from JSPS KAKENHI Grant Numbers 19K23437 and 20K14464. Analysis and data reduction was completed on the Texas Advanced Computing Center's Stampede2 cluster through computing grant TG-AST190001 managed by the XSEDE/ACCESS programs.

\bibliography{main}
\bibliographystyle{aasjournal}

\section{Appendix}
\subsection{Loss Plots}
\label{loss}

\begin{figure*}[htp]
    \centering
    \includegraphics[width=13cm]{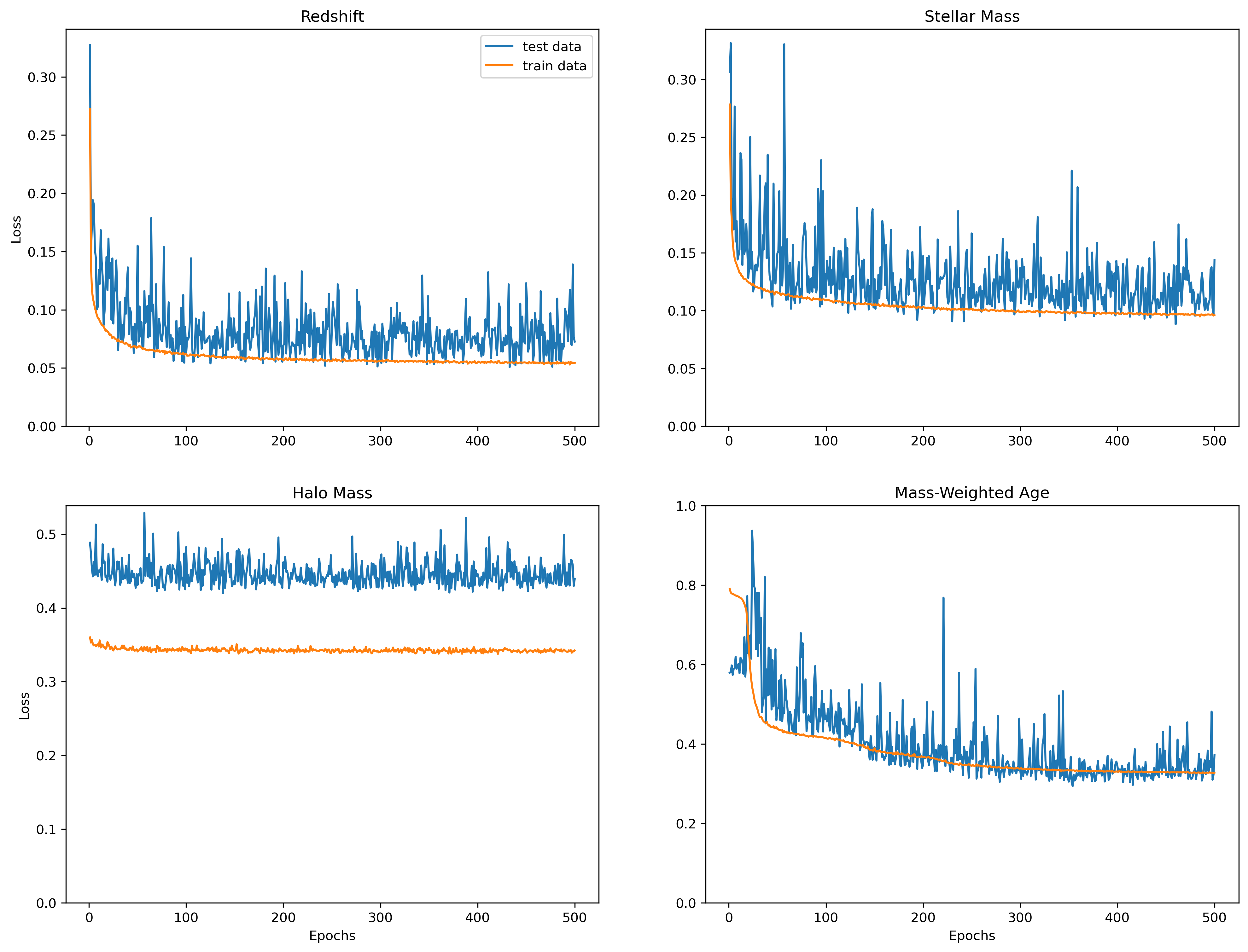}
    \caption{Loss as a function of the number of epochs for each variable in the unconstrained best case. Other than halo mass, which is not predictable by the NNs, there is an overall downward trend in the loss for all variables. Even though there are several spikes in the test loss, they also tend to decrease as the number of epochs increases. This behavior is also observed in the loss plots of the other cases.}
    \label{lossplot}
\end{figure*}

\subsection{Activation Functions}
\label{actv}
The following equations are definitions of the less common activation functions used in this work. The parameter $\alpha$ can be determined by the user, but we use the default $\alpha = 1$ on {\sc Tensorflow}. The constants $\beta$ and $\gamma$ are equal to $1.05070098$ and $1.67326324$, respectively. 

\begin{equation}
    \label{eq:elu}
    \rm{ELU}(x) = 
    \begin{cases}
        x & \text{if } x > 0\\
        \alpha(e^x - 1) & \text{if } x < 0
    \end{cases}
\end{equation}

\begin{equation}
    \label{eq:gelu}
   \rm{GELU}(x) = \frac{1}{2}x(1 + \rm{erf}(x / \sqrt{2}))
\end{equation}

\begin{equation}
    \label{eq:selu}
    \rm{SELU}(x) = 
    \begin{cases}
        \beta x & \text{if } x > 0\\
        \beta\gamma(e^x - 1) & \text{if } x < 0
    \end{cases}
\end{equation}

\begin{equation}
    \label{eq:swish}
    \rm{SWISH}(x) = x * \rm{sigmoid}(x)
\end{equation}

\begin{equation}
    \label{eq:softplus}
    \rm{SOFTPLUS}(x) = \rm{log}(e^x + 1)
\end{equation}

\begin{equation}
    \label{eq:softsign}
    \rm{SOFTSIGN}(x) = x / (|x| + 1)
\end{equation}

\end{document}